# An Exploration of Physiological Responses to the Native American Flute

Eric B. Miller[†] and Clinton F. Goss[‡]

[†]Montclair State University, Montclair, New Jersey; Email: millerer@montclair.edu
[‡]Westport, Connecticut; Email: clint@goss.com

**ABSTRACT**

This pilot study explored physiological responses to playing and listening to the Native American flute. Autonomic, electroencephalographic (EEG), and heart rate variability (HRV) metrics were recorded while participants ($N = 15$) played flutes and listened to several styles of music. Flute playing was accompanied by an 84% increase in HRV ($p < .001$). EEG theta (4–8 Hz) activity increased while playing flutes ($p = .007$) and alpha (8–12 Hz) increased while playing lower-pitched flutes ($p = .009$). Increase in alpha from baseline to the flute playing conditions strongly correlated with experience playing Native American flutes ($r = +.700$). Wide-band beta (12–25 Hz) decreased from the silence conditions when listening to solo Native American flute music ($p = .013$). The findings of increased HRV, increasing slow-wave rhythms, and decreased beta support the hypothesis that Native American flutes, particularly those with lower pitches, may have a role in music therapy contexts. We conclude that the Native American flute may merit a more prominent role in music therapy and that a study of the effects of flute playing on clinical conditions, such as post-traumatic stress disorder (PTSD), asthma, chronic obstructive pulmonary disease (COPD), hypertension, anxiety, and major depressive disorder, is warranted.

## Introduction

The Native American flute, a traditional ethnic wind instrument developed by indigenous Native American cultures, is enjoying a renaissance in various sectors of society. The instrument evolved from traditional uses in courtship (Black Hawk & Patterson, 1834; Burton, 1909), treatment of the sick (Densmore, 1936), ceremony (Gilman, 1908; Stacey, 1906), signaling (Densmore, 1929), legends (Deloria & Brandon, 1961; Densmore, 1923; Erdoes & Goble, 1976; Erdoes & Ortiz, 1984; Wissler, 1905), and as work songs (Densmore, 1957; Winship, 1896).

The design of the Native American flute is "*a front-held, open-holed whistle, with an external block and internal wall that separates a mouth chamber from a resonating chamber*" (R. Carlos Nakai, personal communication, June 21, 2002, as cited in Goss, 2011). The instrument first appeared in the historical record in the early 19th century, and has been known by various names such as "courting flute", "love flute", "plains flute", "woodlands flute", and "śi'yotaŋka" (Densmore, 1918).

The Native American flute is classified in the same family as the recorder.[1] It uses a duct or *flue* to direct the player's airstream, allowing the instrument to be played without the need for players to learn to form an embouchure with their lips. It is distinguished from the recorder by the inclusion of a slow air chamber which precedes the flue, providing an air reservoir that acts as a modest pressure bladder, tending to smooth out changes in breath pressure. Another distinguishing characteristic is its limited pitch range – typically no more than 1.3 octaves from the lowest note on the instrument.

Figure 1 shows the typical elements used in the design of a Native American flute. Since there are no common design standards, contemporary instrument makers take far more freedom in their designs than makers of orchestral wind instruments.

---

[1] In the widely-used classification system of Hornbostel & Sachs (1914).

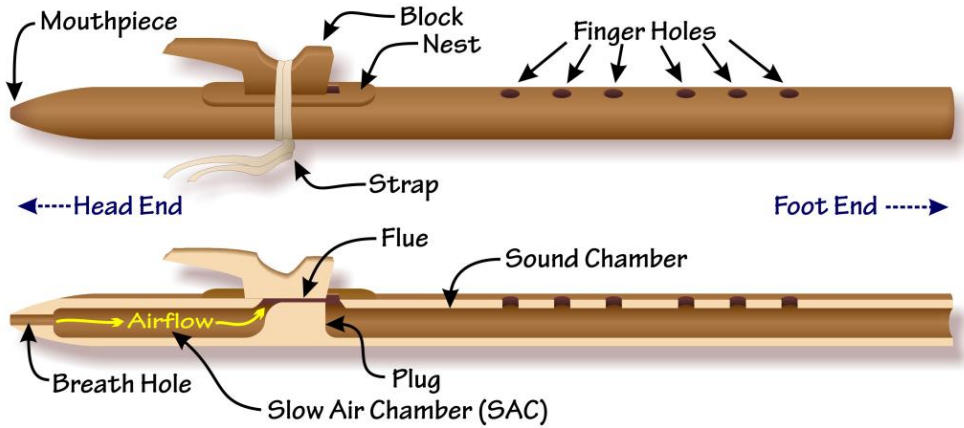

**Figure 1**. Anatomy of the Native American flute.

### Historical Use in Healing

Various forms of flutes and reed aerophones have been used in healing contexts since at least the time of Aristotle (323–373 BCE: Meymandi, 2009) and possibly as far back as the Third Dynasty of Ur in Ancient Mesopotamia (2100–1900 BCE: Krispijn, 2008).

A tradition of *flûtes sacrées* (*flautas sagradas* or *sacred flutes*) is found in a number of indigenous South American cultures (Menezes Bastos & Rodgers, 2007; Piedade, 2004, 2006). The Jesuit priest, José Gumilla (1741), provided an early description of rituals with these instruments and related them to funeral rites held by the Saliva of the Orinoco basin, in present-day Venezuela (Hill & Chaumeil, 2011).

In North America, indigenous rim-blown flute designs have been depicted in religious music or as part of a magical rite (Renaud, 1926). Music played on these flutes was used for Hopi religious ceremonies and during medicine preparation (Hough, 1918).

### Wind Instruments in Healing and Therapy

In 1956, after observing that children with asthma who played wind instruments often did exceptionally well in sports activities, Marks (1974) developed a program using brass instruments, which demonstrated improved lung function parameters. This work inspired the development of the long-running Léčivá Píšťalka [Active Flute] program (Komárková, 2012) in the present-day Czech Republic that uses the soprano recorder for children with asthma. The program, presently known as Veselé Pískání – Zdravé Dýchání [Merry Whistling – Healthy Breathing] (Žilka, 1993), has shown significant improvement in respiratory parameters, posture, and breathing coordination after two years of daily wind instrument playing (Petrů, Carbolová, & Kloc, 1993).

In other studies, Puhan et al. (2006) found that regular playing of the didgeridoo was an effective treatment for patients with moderate obstructive sleep apnea syndrome. The didgeridoo was also found to improve respiratory function and self-reports of health in Aboriginal junior-school and senior-school boys with asthma (Eley & Gorman, 2008). Lucia (1994) found a reduction in panic-fear responses and mood changes in teenage wind instrument players with asthma versus their peers who did not play wind instruments.

A number of studies have investigated the relationship between heart rate variability (HRV) and playing musical instruments with respect to performance anxiety and emotion (Harmat & Theorell, 2010; Harmat et al., 2011; Thurber, 2006; Nakahara, Furuya, Obata, Masuko, & Kinoshita, 2009). However, none have involved the Native American flute.

### Native American Flutes in Healing and Music Therapy

The Native American flute has a reputation for a meditative and healing sound that is compatible with New Age music, and is often heard in meditation centers, museum shops, and yoga studios. This instrument is also flourishing in the expanding social phenomenon of *flute circles* – informal social music gatherings that support the use of the instrument by players with little or no formal music training (Jones, 2010).

In present-day healing and therapy contexts, the Native American flute was reportedly used in hospice care by a music therapist for a Navajo woman (Metzger, 2006), to meet the emotional and spiritual needs of Aboriginal students (Dubé, 2007), and in the treatment of anxiety in individuals diagnosed with a trauma-related disorder (Wiand, 2001, 2006). Aside from these references, a literature search of the ProQuest, JSTOR, and PsychInfo systems yielded no results, demonstrating a paucity of scientific literature seeking to understand the underlying mechanisms behind the cultural tradition that accompanies these instruments.

While Wiand (2001, 2006) was interested in the effect of the Native American flute on anxiety, her measure was by self-report rather than physiological measurement. Her conclusion that Native American flute music appears helpful in treating trauma-related disorder populations is consistent with the popular notion of the Native American flute as a healing instrument. However, her study also highlights the need for validation via objective physiological measures.

*The Present Study*

This pilot study explored physiological responses to the Native American flute. This line of inquiry could assist music therapists using the Native American flute in their practices through increased understanding of potential clinical applications of the Native American flute. We monitored electroencephalographic (EEG) brainwaves, heart rate (HR), electrodermographic activity (EDG), and blood volume pulse (BVP). In post-analysis, derivative measures of HRV were also examined.

Four hypotheses were proposed:

1. Listening to Native American flute music will entrain a meditative brain state, discernible in brainwave patterns of increased EEG alpha and theta with reduced beta activity.

2. Listening to Native American flute music will induce a relaxed state, discernible by autonomic measures of reduced HR, EDG, and electromyogenic activity (EMG), with increases in BVP and HRV.

3. Playing Native American flute will entrain a meditative brain state, discernible in brainwave patterns of increased EEG alpha and theta with reduced beta activity.

4. Playing Native American flutes will induce a relaxed state, discernible by autonomic measures of reduced HR with increases in BVP and HRV.

This investigation also explored differences in the effects of playing a lower-pitched Native American flute versus a higher-pitched Native American flute and differences in the effects of listening to several different styles of flute music as compared to sitting in silence.

**Method**

A convenience sample of 15 Native American flute players was taken from volunteer participants at the 2009 Flute Haven Native Flute School. These flute players did not necessarily have Native American heritage. Participants signed an informed consent and agreed to participate in and of their own free will with an understanding that they could withdraw at any time. Participant data was de-linked from identifying information to protect confidentiality.

Participants varied in gender and in their amount of experience playing Native American flutes.

Participants were asked to bring two of their own Native American flutes to the study: A *lower-pitched flute* with a lowest attainable pitch in the range $A_3$–$E_4$ (220.0–329.6 Hz)[2] and a *higher-pitched flute* with a lowest attainable pitch in the range $G_4$–$E_5$ (329.0–659.3 Hz).

Participants were fitted with a non-invasive EEG sensor at Cz, ear clips for reference and ground, and a finger sensor for autonomic measures on a non-playing finger. All participants listened to the same sequence of instructions, silence, and periods of music on closed-cell headphones.

*Study Outline*

Following pre-recorded instructions, the study conditions comprised:

1. Baseline silence.

2. Listening to solo flute. "Canyon People" (Nakai, 1993, track 7) consists of solo Native American flute in *parlando* style – with no meter or definitive rhythm.

3. Interim silence 1.

4. Playing lower-pitched flute. Participants were asked to play their lower-pitched flute.

5. Listening to rhythmic flute. "Lost" (Ball, 2002, track 6) is a highly rhythmic piano pattern with a melody played on a Native American flute.

6. Interim silence 2.

7. Playing higher-pitched flute. Participants were asked to play their higher-pitched flute.

8. Listening to melodic cello. "Prayer for Compassion" (Darling, 2009, track 2) is a polyphonic, melodic composition containing numerous layers of cello.

9. Post-baseline silence.

Conditions were approximately two minutes in length, except for the two shorter interim silence periods.

*Autonomic Metrics*

Autonomic metrics were sampled at 24 Hz using a MindDrive™ finger sensor (Discovogue Infotronics, Modena, Italy). The finger sensor contained both an electrodermal and a photoplethysmographic (PPG) biosensor. The PPG biosensor measures light transmission through tissue and provides a relative measure of instantaneous peripheral blood volume. Once these readings are calibrated around a zero axis, crossings from negative to positive readings were identified as upward zero crossings. The timespan of a pulsebeat was the time between two neighboring upward zero crossings. Processed data for HR, BVP, and EDG were recorded at one-second intervals.

**Electrodermographic readings**. EDG was taken via finger sensor that provided a measure of electrical conductivity of the skin. Eccrine glands produce minute moisturized particles (sweat) which increase skin conductivity. Eccrine gland output increases with nervous system activation. Galvanic skin response (GSR), measures skin resistance and is an inverse indicator of EDG. Both EDG and GSR measures of electrodermal activity have been used extensively in psychological research and are common measures of autonomic nervous system activity (Andreassi, 2006; Mendes, 2009).

---

[2] The note names in this paper are based on Young (1939).

**Blood volume pulse**. BVP is a relative measure of peripheral blood volume.

**Heart rate**. The HR metric consisted of PPG-recorded pulsebeats in beats/min.

**Heart rate variability**. HRV is the fluctuation in time intervals between adjacent cardiac cycles. Maximum HR minus minimum HR was calculated within estimated breath cycles of 10 (EBC10) and 16 seconds[3] (EBC16) averaged over discrete consecutive windows of those time periods. Both EBC metrics have been shown to correlate very strongly with SDNN – the standard deviation (SD) of intervals between R peaks in adjacent cardiac cycles from normalized ECG data (Goss & Miller, 2013). ECG-derived measures of HRV are the "gold standard" of HRV measurement (Selvaraj, Jaryal, Santhosh, Deepak, & Anand, 2008). Russoniello, Zhirnov, Pougatchev, & Gribkov (2013) showed very strong correlations between HRV metrics from PPG and ECG data.

*EEG Metrics*

Readings from a monopolar EEG sensor placed at Cz were acquired through a BrainMaster™ 2E system (BrainMaster Technologies, Bedford, OH), taken at 256 Hz. Particular attention was given to these EEG bandwidths:[4]

- Delta (0.5–4 Hz), dominates EEG spectral activity in adults during slow-wave sleep, a phase of deep, non-rapid eye movement sleep (Silber et al., 2007);
- Theta (4–8 Hz), usually found during creative processes and deep meditation (Gruzelier, 2009; Miller, 2011; Wright, 2006), as well as working memory (Vernon, 2005) and drowsiness and inattention (Gruzelier & Egner, 2005);
- Alpha (8–12 Hz), associated with a light meditative state (Gruzelier, 2009; Miller, 2011; Wright, 2006), appearing with closing of the eyes and with relaxation, and attenuating with eye opening or mental exertion (Jensen et al., 2005; Lucking, Creutzfeldt, & Heinemann, 1970; Yang, Cai, Liu, & Qin, 2008);
- Sensorimotor rhythm (SMR) (12–15 Hz), inversely related to motor activity or motor imagery (Gruzelier & Egner, 2005; Lubar & Shouse, 1976; Monastra et al., 2005) and positively associated with semantic working memory (Vernon, 2005);
- Beta (15–25 Hz), usually associated with alert, active cognition or anxious concentration (Gruzelier & Egner, 2005; Miller, 2011); and
- Gamma (35–45 Hz), thought to represent neuronal activity that links several distant areas of the brain in a single function (van Deursen, Vuurman, Verhey, van Kranen-Mastenbroek, & Riedel, 2008; Wright, 2006).

In some cases, we also analyzed wide-band beta in the frequency range 12–25 Hz as well as theta and alpha in the range 4–12 Hz. We also monitored muscle movement artifact using an algorithm that processed EEG activity in the 25–35 Hz band.[5] This provided a convenient indication of potential EEG signal contamination.

*Analysis*

Following acquisition, data artifacts were identified with a software algorithm based on the standard score for each data point. Data points exceeding ±4σ were then removed during a visual inspection phase.

Participants were categorized for analysis by experience, gender, and age:

- novice (less than 3 years of experience playing Native American flutes, $n = 7$) versus experienced ($n = 8$) flute players;
- male ($n = 10$) versus female ($n = 5$); and
- younger (less than the median age of 58.4 years, $n = 8$) versus older ($n = 7$).

---

[3] The length of estimated breath cycles while playing flute are supported by an informal breathing-rate survey conducted concurrently with this study. Native American flute players ($N = 28$, mean age = 59.9 years) self-assessed the number of inhalations taken during one minute of "normal or average playing" on various flutes. Their reports average 10.31 seconds per breath cycle ($SD = 4.51$). Compared with a normal respiratory cycle of 3–5 seconds in adults (Lindh, Pooler, Tamparo, & Dahl, 2009), flute players tend to extend their breath cycles while playing.

[4] The characteristics described for the EEG bands are a simplification provided for general background and do not reflect the diversity of associated functional states and neural communications (Gruzelier & Egner, 2005).

[5] The algorithm for deriving EMG from EEG data was used in the Lexicor NRS–2D series of neurofeedback training machines. While those particular machines shipped with a default software setting that designated amplitudes of 25–32 Hz over 15 µV as EMG artifact, a study of attention deficit/hyperactivity disorder conducted by the City of Philadelphia Office for Mental Health (Berman, 2001) used a slightly more conservative bandwidth: 25–35 Hz averaged over 250 ms to detect more muscle activity (Marvin Berman, personal communication, April 8, 2000). The premise is that, while EMG artifact cuts across the entire spectrum of 0–100 Hz, a representative sample may be acquired from just above the Beta range.

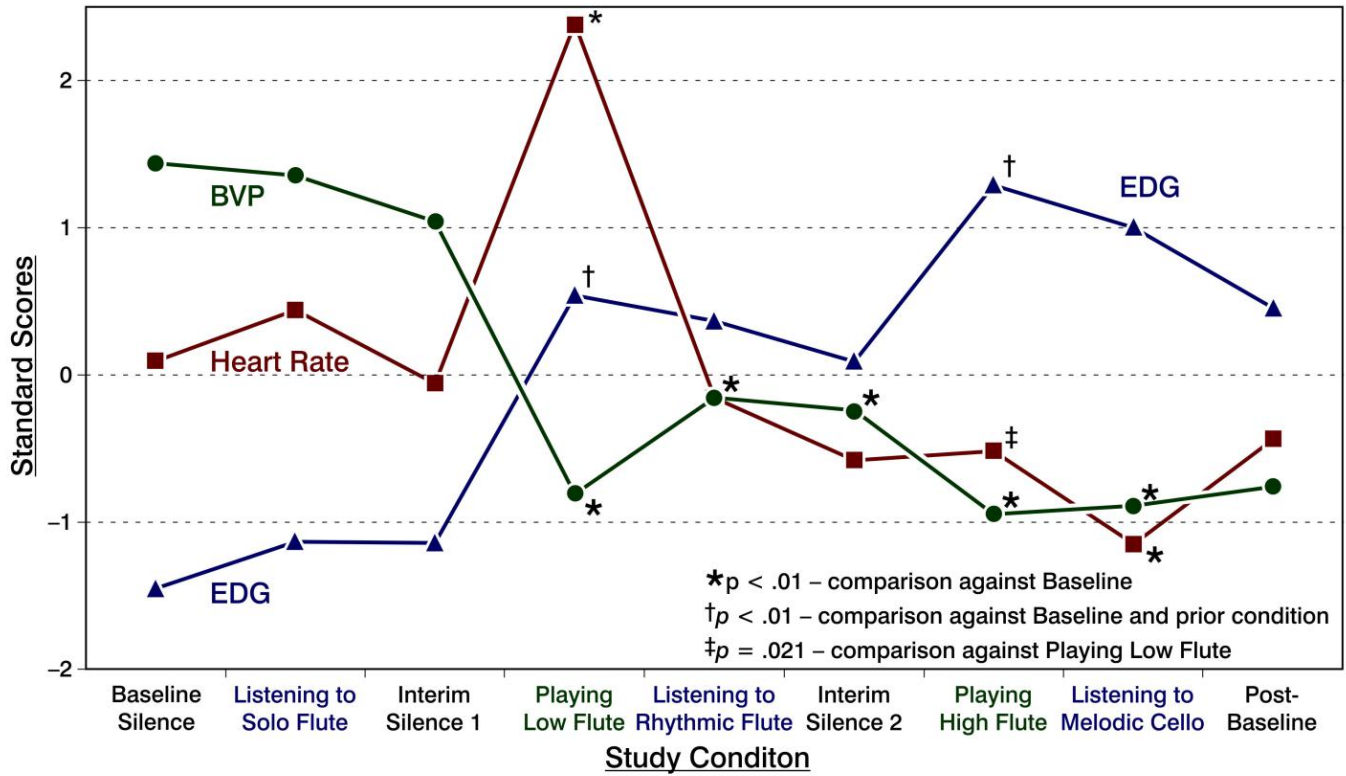

**Figure 2**. Autonomic measures for the full sample (*N* = 15) across all study conditions. BVP = blood volume pulse. EDG = electrodermographic readings. †indicates a significance of *p* < .01 when compared both with the baseline condition and the immediately preceding condition.

### Results

For this pilot study, data analysis examined a wide range of measures and possible outcomes. Because multiple statistical inferences were considered simultaneously, the statistical measures presented should be considered exploratory.

These repeated-measures ANOVAs were calculated: theta and alpha: $F(6, 66) = 5.368$, $p < .001$, $\eta_p^2 = .328$; wide-band beta: $F(6, 66) = 3.280$, $p = .007$, $\eta_p^2 = .230$; EBC16: $F(6, 66) = 2.828$, $p = .016$, $\eta_p^2 = .205$; EBC10: $F(6, 66) = 3.159$, $p = .009$, $\eta_p^2 = .223$.

*Autonomic Response*

Figure 2 plots the trends in autonomic responses to the study conditions. A generally inverse relationship between BVP and EDG can be seen. This pattern helps to corroborate the validity of the measures since it would be expected that, during the passive silence and listening conditions, sympathetic nervous system activity would decrease, allowing for increased blood flow.

As expected, EDG increased from baseline to both playing lower-pitched flute ($p < .001$) and playing higher-pitched flute ($p = .001$). However, the trend of EDG in Figure 2 shows the slow decay that is often seen when recovering from quick onset of activation (Miller, 2011).

HR increased from baseline to playing lower-pitched flute ($p < .001$). This increase in physical arousal was corroborated by a corresponding increase in EDG ($p < .001$) and a decrease in BVP from baseline to playing lower-pitched flute ($p < .001$).

Mean HR, however, decreased from baseline to playing higher-pitched flutes. HR was significantly lower when playing higher-pitched flutes than lower-pitched flutes ($p = .021$). We speculate that this may be due to an ordering effect, where playing the flute for the first time caused some anxiety that was not present in the second flute playing condition. This decrease in HR with a concurrent increase in EDG indicates differing reactions between the vagal response and the exocrine system response – a divergence that begs further investigation.

When listening to the melodic cello music, participants displayed lower HR from baseline ($p = .008$). This HR decrease did not occur in the other two listening conditions.

*EEG Response*

Figure 3 plots the trends in EEG response to the study conditions.

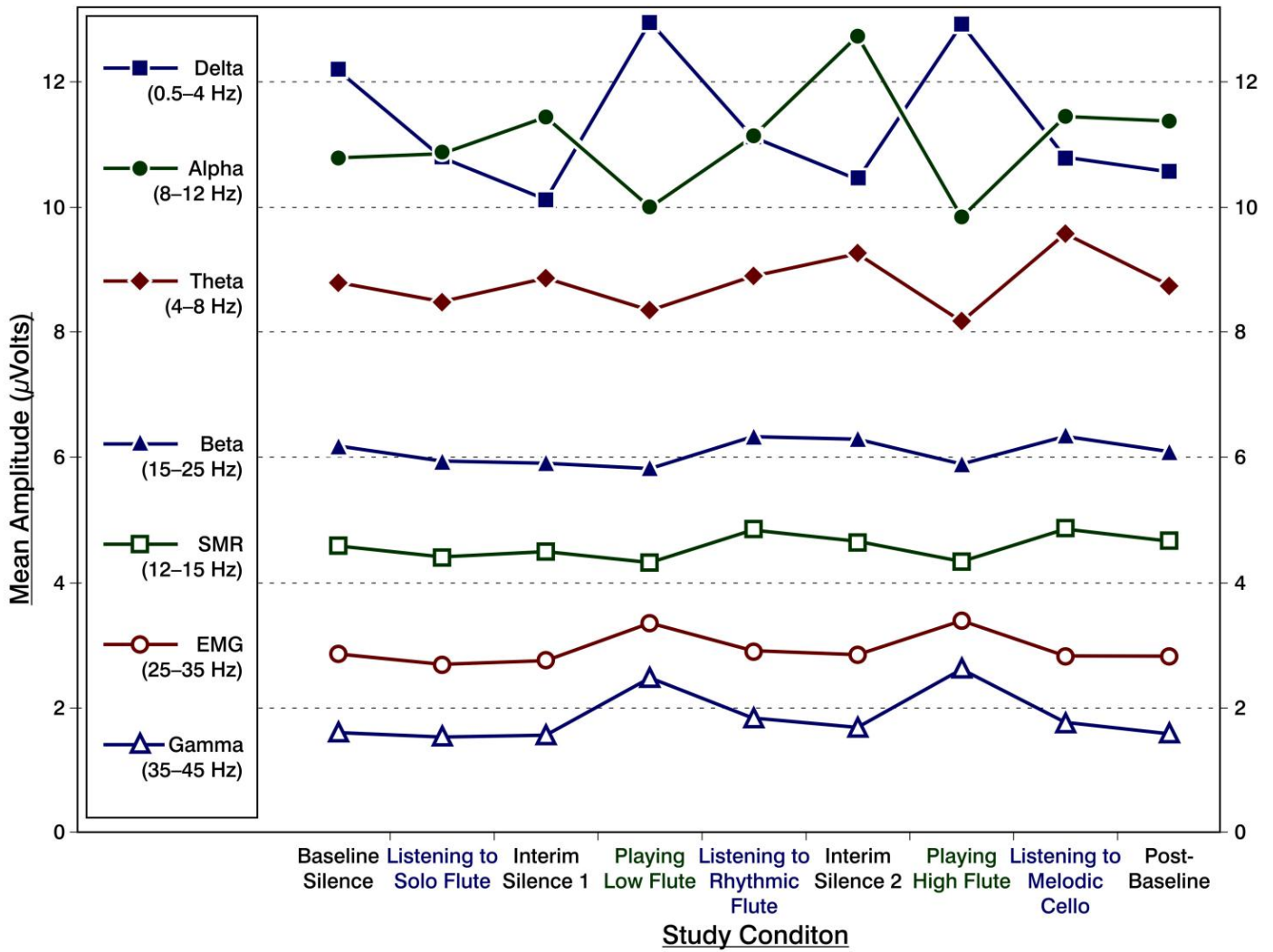

**Figure 3**. EEG activity for the full sample (*N* = 15) across all study conditions.

Figure 3 is remarkable for the predominantly inverse and reversible relationship between the delta and alpha bands. The general trend of the theta band follows the alpha band. However, contrary to our expectations, mean alpha band activity was reduced for both playing conditions.

Activity in the SMR and beta bands was reduced during the two flute playing conditions. While these trends were not statistically significant, the reduced activity in these bands while playing indicates that the physical movements involved with flute playing did not mask an actual decrease in alpha and theta during the flute playing conditions.

Mean delta band activity decreased from baseline for all three listening conditions. This decrease was significant for listening to melodic cello ($p = .032$) and approached significance for listening to rhythmic flute ($p = .057$).

No significant differences were found for theta band activity for the full sample across conditions as compared with baseline.

Alpha band activity was highest across the study conditions during interim silence 2 – significantly higher than the two preceding conditions, playing lower-pitched flute ($p = .006$) and listening to rhythmic flute ($p = .008$), as well as the aggregate of the listening conditions ($p = .010$) and the aggregate of the playing conditions ($p = .004$). Mean alpha band activity during interim silence 2 was also higher than during baseline silence, a result that approached significance ($p = .067$).

Figure 4 shows differences in EEG beta response among the listening conditions. When compared with listening to solo flute, beta response was significantly higher when listening to melodic cello ($p = .039$), as well as listening to rhythmic flute ($p = .029$).

We hypothesize that the rhythmic structure played a role in these significant differences, possibly indicating a reduction in mental activity and cognitive tasking while listening to music with less rhythmic structure.

Beta band activity showed no significant differences from baseline silence. However, wide-band beta response when listening to solo flute was lower than the aggregate of the silence conditions ($p = .013$).

As expected, EMG activity increased from baseline to playing lower-pitched flute ($p = .021$). We surmise that the increase in EMG resulted from the voluntary muscle movements involved in the playing conditions.

EMG increase was not significant from baseline to playing higher-pitched flute. We surmise that this was due to the tendency for higher-pitched flutes to be smaller, lighter, and to involve less finger spread. EMG from

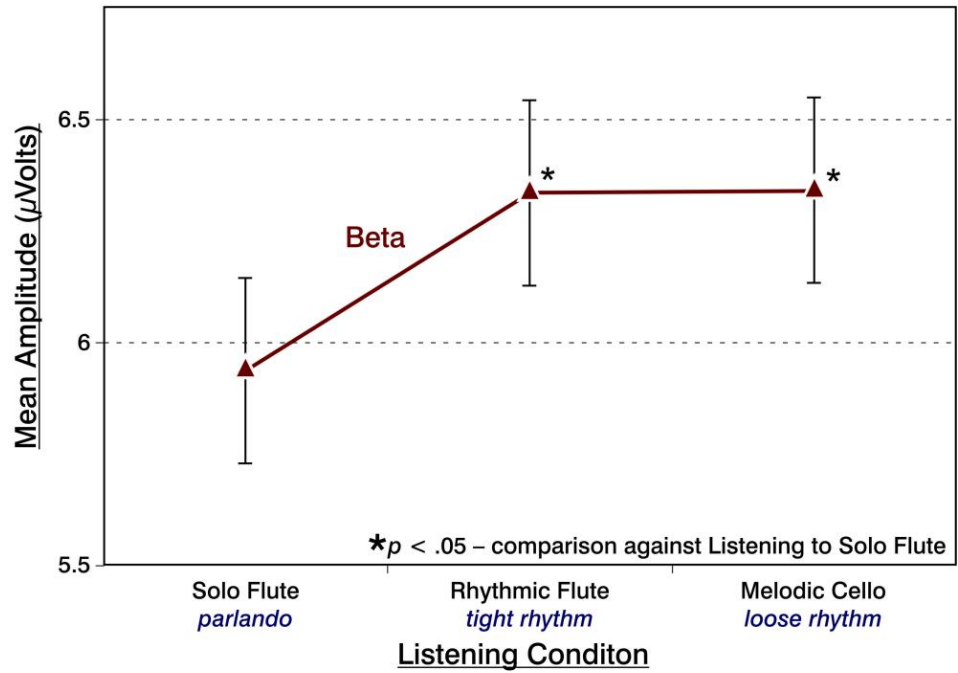

**Figure 4**. Beta (15–25 Hz) activity for the full sample ($N = 15$) across the listening conditions. The error bars depict ±1 standard deviation of that EEG band across all study conditions.

**Table 1.** Metrics for Two Exemplary Participants showing Alpha Suppression and Enhancement

| Participant Type Condition | Alpha | | | Alpha / Beta Ratio | | |
|---|---|---|---|---|---|---|
| | M | p-Base | p-Post | M | p-Base | p-Post |
| **Alpha Suppression Exemplary Participant** | | | | | | |
| Baseline | 16.34 | | .009 | 3.29 | | .028 |
| Playing lower-pitched flute | 11.54 | < .001 | < .001 | 2.16 | < .001 | < .001 |
| Playing higher-pitched flute | 10.72 | < .001 | < .001 | 1.75 | < .001 | < .001 |
| Post-baseline | 19.09 | .009 | | 3.82 | . 028 | |
| **Alpha Enhancement Exemplary Participant** | | | | | | |
| Baseline | 10.94 | | < .001 | 1.73 | | < .001 |
| Playing lower-pitched flute | 18.68 | < .001 | .130 | 2.42 | < .001 | .731 |
| Playing higher-pitched flute | 19.14 | < .001 | .053 | 2.54 | < .001 | .221 |
| Post-baseline | 16.96 | < .001 | | 2.37 | < .001 | |

*Note*: Alpha band activity is in μVolts. *p*-Base = comparison against baseline silence. *p*-Post = comparison against post-baseline silence. Single-subject Student's t-Tests are two-tailed heteroscedastic comparisons.

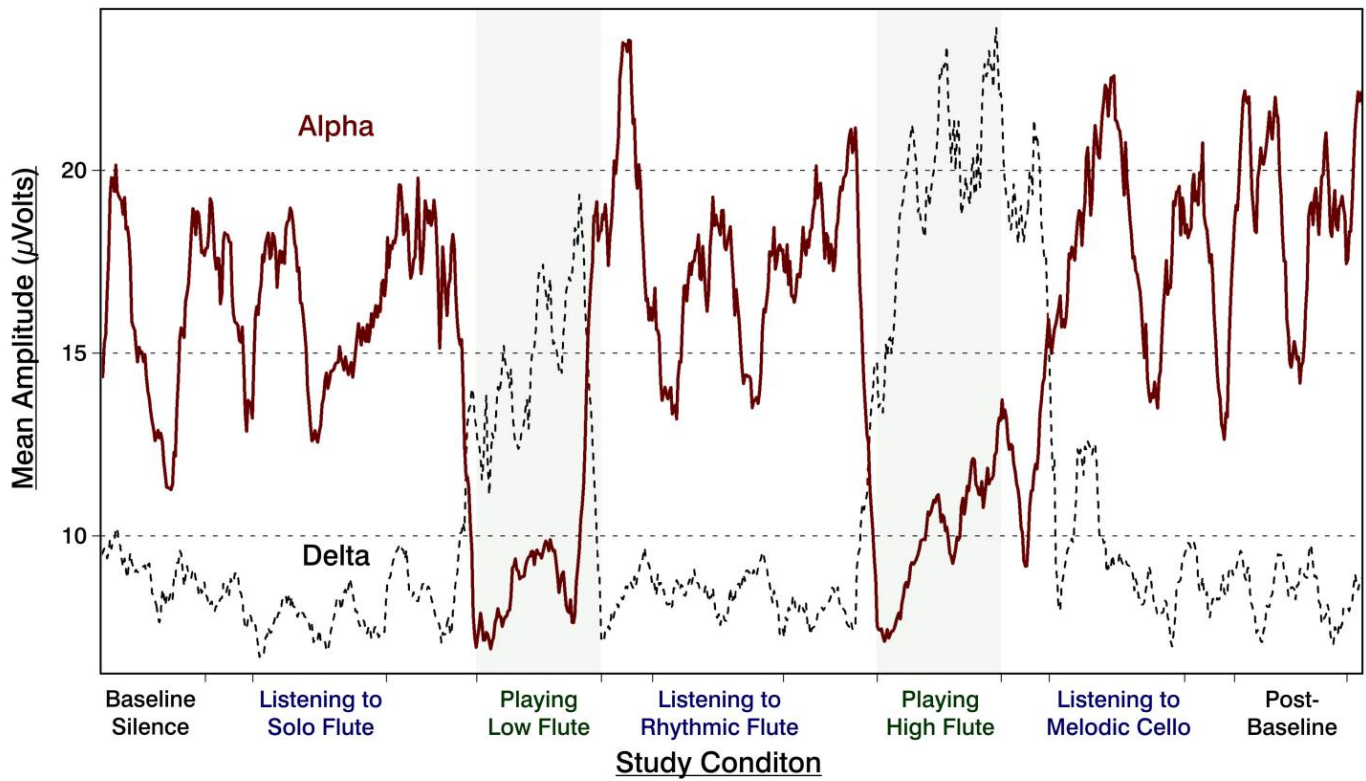

**Figure 5**. EEG alpha (8–12 Hz) and delta (0.5–4 Hz) activity for a participant showing alpha suppression and alpha–delta crossover.

baseline silence to all non-playing conditions shows no significant differences, with the mean EMG showing a decrease in some cases.

### EEG Trends in the Novice Subgroup

The novice subgroup ($n = 7$) demonstrated some significant EEG responses to the flute playing conditions that were not evident in the full sample:

- Alpha response decreased from baseline to playing higher-pitched flute ($p = .030$) and from the prior interim silence to playing higher-pitched flute ($p = .019$).
- Theta increased from baseline to post-baseline ($p = .031$).
- Beta response decreased from baseline to both playing lower-pitched flute ($p = .036$) and playing higher-pitched flute ($p = .018$), as well as the aggregate of the two flute playing conditions ($p = .021$).

In contrast to the decrease in beta during the playing conditions seen in the novice subgroup, the experienced subgroup showed a slight increase in mean beta response from baseline to playing lower-pitched flute, playing higher-pitched flute, and the aggregate of the two flute playing conditions. This contrast may indicate a differential reduction in mental activity when playing Native American flutes based on the level of experience with the instrument.

### Alpha Suppression and Alpha Enhancement

Some individual participants showed a distinct pattern of lowered alpha band response during the flute playing conditions, while other participants showed the opposite effect of increased alpha band response during flute playing.

Figure 5 shows one participant that exemplifies the alpha suppression pattern – remarkable for the dramatic reduction in alpha activity that appeared only during the two flute playing conditions. As shown in Table 1, this participant showed significantly lower alpha as well as alpha/beta ratios when comparing lower-pitched and higher-pitched flute playing to baseline and post-baseline silence.

Table 1 also shows an exemplary participant exhibiting alpha enhancement, with significantly higher alpha and alpha/beta ratios for both lower-pitched and higher-pitched flutes when compared to baseline. In addition, significantly higher alpha activity from baseline to post-baseline was shown in both the alpha suppression and alpha enhancement exemplary participants.

### Alpha Correlation with Experience Playing

To analyze correlations, we defined Δalpha for a study condition to be the measure of alpha activity in that study condition divided by alpha activity during baseline silence.

**Table 2.** Trends between the First Half and the Second Half of the Listening Conditions

| Metric | Listening to … | | | | | |
| --- | --- | --- | --- | --- | --- | --- |
| | Solo Flute | | Rhythmic Flute | | Melodic Cello | |
| | Δ | p | Δ | p | Δ | p |
| HR | −0.3% | .604 | −1.1% | .210 | +0.6% | .479 |
| BVP | −8.2% | .115 | +0.4% | .950 | +11.4% | .297 |
| EDG | +0.3% | .777 | −2.0% | .045 * | −2.6% | .011 * |
| Delta | +3.5% | .334 | +0.5% | .900 | −9.0% | .013 * |
| Theta | −0.7% | .787 | +2.7% | .161 | −6.1% | .232 |
| Alpha | +6.6% | .040 * | +1.6% | .651 | +5.1% | .266 |
| SMR | +5.1% | .006 * | +1.9% | .459 | −0.6% | .834 |
| Beta | −0.5% | .819 | +3.0% | .223 | −2.2% | .388 |
| Gamma | +7.4% | .123 | −5.3% | .328 | −8.7% | .003 * |
| Alpha / Beta ratio | +6.7% | .020 * | −0.6% | .884 | +7.5% | .035 * |
| Alpha / Theta ratio | +8.3% | .025 * | −2.3% | .643 | +8.4% | .023 * |

*Note*: Δ = change between the first half and the second half of the study condition. HR = heart rate. BVP = blood volume pulse. EDG = electrodermographic readings. SMR = sensorimotor rhythm. * = $p < .05$.

For the full sample, the number of years of experience reported by the participants playing Native American flutes was strongly correlated to Δalpha for playing lower-pitched flute ($r = +.699$) and for playing higher-pitched flute ($r = +.691$), as well as the aggregate of the two flute playing conditions ($r = +.700$).

### Trends during Study Conditions

A number of participants demonstrated noticeable trends within various study conditions. To analyze these trends, we divided each study condition into two time periods of equal length.

**Trends during baseline silence**. There were no significant trends in any of the participants for HR, EDG, BVP, or EMG during the baseline silence condition. EEG activity showed no significant trends for delta, alpha, SMR, beta, or gamma. However, there was a decrease in theta activity ($p = .046$) from the first half to the second half of the baseline silence.

**Trends during listening conditions**. Table 2 shows changes in autonomic and narrow EEG band metrics from the first half to the second half of each listening condition. In contrast with baseline silence, two listening conditions show modest, but statistically significant, decreases in EDG, suggesting a relaxing effect after initial activation of the autonomic nervous system.

**Table 3.** Trends between the First Half and the Second Half of the Playing Conditions

| Metric | Playing Lower-pitched Flute | | Playing Higher-pitched Flute | |
| --- | --- | --- | --- | --- |
| | Δ | p | Δ | p |
| HR | −1.3% | .446 | −1.7% | .215 |
| BVP | +26.3% | .005 * | +16.1% | .034 * |
| EDG | +1.0% | .514 | −2.4% | .046 * |
| Delta | −1.9% | .696 | +2.0% | .536 |
| Theta | +8.5% | .052 | +8.8% | < .001 * |
| Alpha | +16.7% | .009 * | +5.9% | .236 |
| SMR | +10.0% | .002 * | +9.2% | .006 * |
| Beta | +7.4% | < .001 * | +9.4% | .002 * |
| Gamma | +0.4% | .962 | +12.5% | .045 * |

*Note*: Δ = change between the first half and the second half of the study condition. HR = heart rate. BVP = blood volume pulse. EDG = electrodermographic readings. SMR = sensorimotor rhythm. * = $p < .05$.

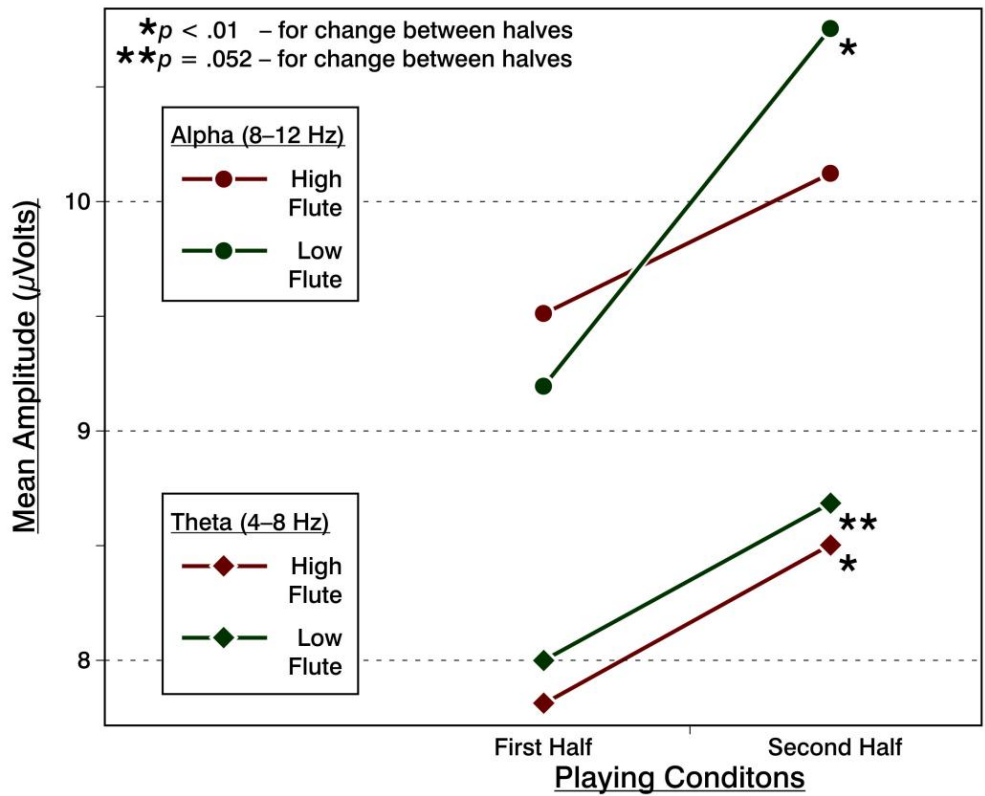

**Figure 6**. Alpha and Theta increase during the Native American flute playing conditions.

Table 2 also includes two notable ratios that show significant increases for two of the listening conditions. Increasing alpha relative to beta suggests a move toward a light meditative state with reduced cognitive tasking.

**Trends during playing conditions**. Table 3 shows changes in autonomic and narrow EEG band metrics from the first half to the second half of each playing condition.

Note the decreasing trend for EDG while playing higher-pitched flute and the significant rise in BVP and slight decrease in mean HR for both playing conditions. These trends suggest that the relaxation response, seen in two of the listening conditions above, is also evident when playing higher-pitched flute.

In the aggregate of the two flute playing conditions, Figure 6 shows a significant increase in theta ($p = .007$). Individually, theta increase is strongly significant playing higher-pitched flute ($p < .001$) and approached significance playing lower-pitched flute ($p = .052$). Playing lower-pitched flute was accompanied by increased alpha ($p = .009$), but alpha increase for higher-pitched flute was not significant ($p = .236$).

*HRV Response*

Figure 7 shows the trends in the two HRV metrics across the study conditions. The two metrics exhibit a strong positive correlation ($r = +.961$).

The specifics on these HRV trends for the playing and listening conditions are shown in Table 4.

HRV increased significantly for the two flute playing conditions from the baseline silence condition. Both HRV metrics also increased for the two playing conditions from their prior interim silence conditions: interim silence 1 to playing lower-pitched

**Table 4.** Comparison of HRV Metrics across the Study Conditions

| Condition | EBC10 | | EBC16 | |
|---|---|---|---|---|
| | *M* | *p* | *M* | *p* |
| Baseline | 1.66 | | 1.95 | |
| Playing … | | | | |
|    Lower-pitched flute | 2.66 | .004 * | 3.39 | .003 * |
|    Higher-pitched flute | 2.94 | < .001 * | 3.71 | < .001 * |
| Listening to … | | | | |
|    Solo flute | 1.62 | .852 | 2.20 | .351 |
|    Rhythmic flute | 2.35 | .082 | 3.22 | .036 * |
|    Melodic cello | 3.15 | < .001 * | 4.35 | < .001 * |
| Post-baseline | 2.48 | .027 * | 3.33 | .006 * |

*Note*: Mean values are in beats/min. EBC16 and EBC10 are metrics of heart rate variability: the average of the differences between the maximum heart rate and the minimum heart rate within discrete consecutive windows of 16 and 10 seconds, respectively. *p* values are for comparisons against the baseline silence condition. * = $p < .05$.

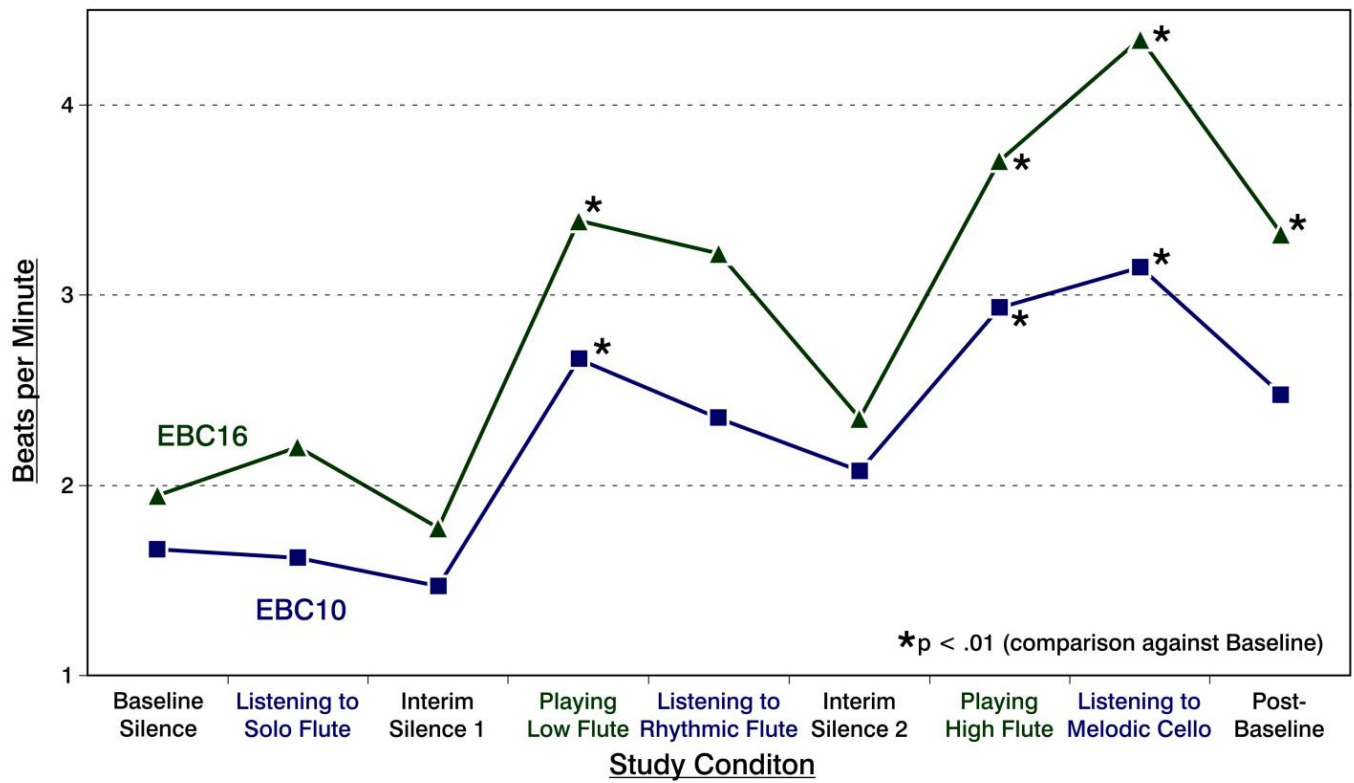

**Figure 7**. Heart rate variability (HRV) metrics for the full sample (*N* = 15) across all study conditions. EBC16 and EBC10 are metrics of HRV: the average of the differences between the maximum heart rate and the minimum heart rate within discrete consecutive windows of 16 and 10 seconds, respectively.

flute for EBC10 ($p = .005$) and for EBC16 ($p = .011$); interim silence 2 to playing higher-pitched flute for EBC10 ($p = .117$) and for EBC16 ($p = .044$).

Listening conditions showed significant increases in HRV metrics from baseline for EBC10 and EBC16 when listening to melodic cello ($p < .001$) and EBC16 when listening to rhythmic flute ($p = .036$). These results may have been related to an ordering effect, since these listening conditions followed playing conditions that showed increased HRV metrics. The increase in both HRV metrics carried into the post-baseline silence, with significant increases from the initial baseline silence condition. The two HRV metrics were very strongly correlated with EDG: EBC10 ($r = +.955$) and EBC16 ($r = +.917$).

To highlight changes in HRV, we define ΔHRV as the measure of HRV in a study condition divided by HRV during baseline silence, expressed as a percentage of change. ΔHRV for the aggregate of the two playing conditions is +78.4% for EBC10 and +89.4% for EBC16. The average ΔHRV for the two metrics was +80.3% ($p = .004$) for playing lower-pitched flute and +87.5% ($p < .001$) for playing higher-pitched flute. A combination of these metrics gave an average ΔHRV across the two EBC metrics and the two playing conditions of +83.9% ($p < .001$).

ΔHRV did not correlate strongly with age, gender, or experience playing Native American flutes. However, participants with less experience at meditation practices showed greater ΔHRV: for the full sample, the number of years of experience that participants reported in a meditation practice was negatively correlated to ΔEBC16 for the aggregate of the playing conditions ($r = -.612$).

### Discussion

### *Results in Light of the Hypotheses*

We established several specific hypotheses regarding listening and playing conditions that could be relevant to clinical music therapy.

**Alpha and theta**. With regard to our expectations of increased alpha during listening and playing conditions, we in fact found the reverse to be the case over the whole sample for flute playing and some flute listening conditions. One possible explanation is that the alpha suppression and alpha enhancement subtype patterns identified had a cancelling effect that masked alpha trends in the full sample. With individual participants showing significant alpha increase, we can envision future research aimed at determining what characteristics contribute to this pattern (such as experience playing the flute or years of meditation practice).

Another possible factor limiting the increase of alpha could be that the time period of two minutes was not a sufficient length of time to achieve a meditative state deep enough to be discernible by EEG rhythms. Trends within the lower-pitched flute playing condition showed increasing theta ($p = .052$) and alpha ($p = .009$), while trends within the higher-pitched flute playing condition showed significantly increasing theta ($p < .001$) but not alpha.

In both cases, with the trend of increasing alpha and theta activity, a longer period of time for the flute playing condition might yield an increase in alpha and or theta sufficient to differentiate flute playing from baseline.

**Beta**. Our hypotheses relating to beta suppression were partially supported in this study, with significantly lower wide-band beta during listening to solo flute than the aggregate silence. Listening to solo flute was also accompanied by lower beta, as shown in Figure 4, when compared with the other two listening conditions.

**Autonomic metrics**. Indications of participants moving toward a relaxed state during the flute playing conditions include increases in BVP and a decrease in EDG during higher-pitched flute playing, without an increase in HR in either condition. We also note a significant decrease in HR from lower-pitched flute playing to higher-pitched flute playing.

While the slow-wave EEG metrics point toward a stronger relaxation effect for playing lower-pitched flutes, the autonomic metrics appear to indicate a stronger relaxation effect for playing higher-pitched flutes. However, we are skeptical of the implications of the autonomic metrics when considering that EDG was highest during the higher-pitched flute playing condition.

*Heart Rate Variability*

HRV (not to be confused with cardiac dysrhythmia) is a characteristic of healthy individuals (Wheat & Larkin, 2010). Low HRV is correlated to a number of medical and psychological diseases, such as anxiety (Friedman, 2007), hypertension (Elliot et al., 2004), chronic obstructive pulmonary disease (COPD) (Giardino, Chan, & Borson, 2004), and depression (Nahshoni et al., 2004). Low HRV is also a prognostic indicator of sudden cardiac death (Goldberger, 1991; Goldberger, Rigney, Mietus, Antman, & Greenwald, 1988).

Several studies have investigated the relationship between HRV and playing musical instruments with respect to performance anxiety (Harmat & Theorell, 2010; Harmat et al., 2011; Nakahara, Furuya, Obata, Masuko, & Kinoshita, 2009; Thurber, 2006). However, none have involved the Native American flute.

We found significant increases in HRV when playing both lower-pitched and higher-pitched Native American flutes. Biofeedback training to raise HRV focuses on breathing techniques, while receiving visual or aural representations of the immediate effects of breathing. We suggest that playing Native American flutes may have an analogous effect to the breath-training component of biofeedback training to raise HRV.

Music playing showed a stronger effect on HRV in this study than music listening. This contrasts with Nakahara, Furuya, Francis, & Kinoshita (2010), who report a decrease in the RMSSD measure of HRV (the root mean square of the successive differences between adjacent R–R intervals from normalized ECG data) from resting to performance by elite pianists during solo performance. Although HRV was highest during the final listening condition, this could have been due to an ordering effect, since it immediately followed playing higher-pitched flute.

*Clinical Implications for Music Therapy*

The finding of increased HRV during the flute playing conditions indicates the potential for use of the Native American flute in treatment of a range of clinical conditions that a music therapist might encounter. Biofeedback training to raise HRV has been found to have various degrees of effectiveness in the treatment of asthma (Lehrer et al., 1997, 2004), hypertension (Elliot et al., 2004), anxiety (Henriques, Keffer, Abrahamson, & Horst, 2011), COPD (Giardino, Chan, & Borson, 2004), post-traumatic stress disorder (PTSD) (Zucker, Samuelson, Muench, Greenberg, & Gevirtz, 2009), and recurrent abdominal pain (Sowder, Gevirtz, Shapiro, & Ebert, 2010). In the treatment of major depressive disorder, biofeedback training to raise HRV demonstrated effects that appeared stronger than most selective serotonin reuptake inhibitors, suggesting that this approach may provide a non-pharmacological alternative treatment method (Karavidas, 2008; Karavidas et al., 2007).

In a review of the literature on biofeedback training to raise HRV, Wheat and Larkin (2010) opined:

> *Significant improvements in clinical outcomes were overwhelmingly evident in the reviewed literature. This is particularly notable given that such changes cut across several disease states … Therefore, HRV BF should be considered seriously as a viable avenue through which to supplement traditional treatments of various illnesses. (p. 237)*

The broad range of applicability of biofeedback training to increase HRV (Wheat & Larkin, 2010) raises a key question for further investigation: Could playing Native American flute prove efficacious for some of these clinical conditions?

We also note that the integration of mental and respiratory functions is central to many Eastern meditation practices, such as Yoga, Qigong, and Zazen (Lehrer, Vaschillo, & Vaschillo, 2000). A central tenet that the mind and breathing are interdependent is embodied in the writings of Yue Yanggui: "*the tranquility of the mind regulates the breathing naturally and, in turn, regulated breathing brings on concentration of the mind naturally*" (Meihua Wen Da Plan [Questions and Answers of Meihua], cited in Xiangcai, 2000, p. 7).

### *Limitations of the Present Study*

The sample size of this pilot study may not have been sufficient to distinguish statistical differences in some of the interesting trends.

The convenience sample of participants ranged in age from 52 to 70 years, with a mean age of 58.4 years. This age bracket is more likely to yield lower HRV readings than a younger population (Moss, 2004; Sandercock et al., 2005).

Playing style during the two playing conditions was not recorded. Players may take full breaths between long passages, shallow breaths for short passages and rhythmic melodies, or even engage in *circular breathing*, which allows continuous playing using small, frequent inhalations through the nose. Although not formally documented, no circular breathing was observed by the investigators. Controlling for a full range of playing styles in future studies is warranted.

The finger sensor may have been bothersome to some participants during the playing conditions. The sensor was placed on the fifth digit of the hand closest to the foot end of the flute – a finger that is not normally used in covering the finger holes of the instrument. However, some players position that finger on the barrel of the instrument for stability. A possible alternative might be to explore the use of ear or toe sensors (Allen, 2007).

The single-sensor EEG interface lacked the ability to localize brain waves. The results could possibly have been more discriminating if a qEEG topographical brain map was employed.

Future analysis could run EEG data through cleaning algorithms, such as those implemented in NeuroGuide software (Thatcher, 2008), to determine if results are consistent.

Distinguishing myogenic from neurogenic signal sources in scalp recordings of high-frequency EEG is known to be problematic (Whitham et al., 2008). Goncharova, McFarland, Vaughan, & Wolpaw (2003) reports a broad frequency distribution of EMG from 0 Hz to greater than 200 Hz with highest amplitudes frontally in the range of 20–30 Hz. Even weak EMG is detectable in scalp recordings in frequencies down into the 8–13 Hz range in some individuals (Shackman et al., 2009).

The time for each study condition – approximately two minutes – is below the recommended five-minute period recommended for the reliable calculation of SDNN metrics for HRV (Task Force, 1996). However, as noted in Berntson et al. (1997), it may not be possible to maintain a stable psychological or cognitive state over a period of five minutes. In any case, it is appropriate to compare SDNN in different periods only if the durations for those study periods are the same (Task Force, 1996).

Although we did not measure breathing frequency of participants during this study, we can infer that the length of an average breath cycle lengthens when playing flute from the normal respiratory cycle of 3–5 seconds in adults (Lindh, Pooler, Tamparo, & Dahl, 2009) to approximately 10 seconds – the mean reported breath cycle in the informal breathing-rate survey. This closely matches the breath cycle of approximately 10 seconds in paced breathing that produces the maximal increases in HRV during biofeedback training to raise HRV (Vaschillo, Vaschillo, & Lehrer, 2006).

While aggregating the silent periods provided a solution in this study, longer periods of silence would allow for stronger comparative analyses.

We did not control for the level of intraoral breath pressure that participants produced, nor the breath volume required during the playing conditions. Breath pressure and breath volume are affected by the resistance provided by the particular instrument, as well as the volume of the sound produced (Goss, 2013).

In this study, participants played their own personal flutes, which increased the variability of breath pressure and breath volume.

No data was collected on the medical conditions, prescribed medications, or other health-related attributes of the participants. In this study, the use of a baseline design controls for the potential effects of medications on EEG measurements across conditions, since we are looking at relative changes from baseline to other study conditions and within condition trends.

Likewise, we did not collect data on the musical preferences of the participants. This could have provided insight into how musical preferences of, for example, parlando style versus rhythmic style, correlate with the physiological responses to listening and playing.

Finally, this study did not control for the sequence of playing and listening conditions that may have contributed to ordering effects.

### *Conclusions and Directions for Future Studies*

This study was prompted by a concordance of factors suggesting that playing the Native American flute and/or listening to Native American flute music may effect a variety of psychological changes with the potential for use in treating clinical conditions in a music therapy context. These factors include: the cultural traditions surrounding the role of the instrument in healing, prior studies and anecdotal evidence of the effectiveness of various wind instruments for a range of clinical conditions, and the widespread availability and increasing use of the instrument in present society. Despite the limitations noted, this study provides an initial investigation of the effects of the Native American flute based on physiological measurements.

While we found some results that support our hypotheses regarding physiological responses to playing and listening to Native American flutes, we have also raised a host of questions. In addition to the specific improvements in study design implied by the Limitations section above and the general directions suggested in the prior Clinical Implications for Music Therapy section, we offer this list of potential questions and directions for future research studies:

- Can differing subtypes based on alpha band response to the flute playing conditions be confirmed and, if so, do they correlate to other factors?
- Are the effects observed in the present study also found in other wind instruments, particularly wind instruments that are outside the domain of orchestral wind instruments that have been the subject of the majority of research studies on wind instruments?
- Can the differing reactions between the vagal response and the exocrine system response implied by our finding of decreased HR with concurrent increased EDG in the playing higher-pitched flute condition be confirmed and, if so, correlated to other factors?

Considering the increased HRV during flute playing found in this study, and the previously demonstrated effect of biofeedback training to raise HRV on a range of clinical conditions, we suggest that the most compelling direction for future research would be a direct investigation of the effect of a music therapy program of Native American flute playing for clinical conditions, such as asthma, COPD, PTSD, recurrent abdominal pain, hypertension, anxiety, fibromyalgia, and major depressive disorder.

## References

Allen, J. (2007). Photoplethysmography and its application in clinical physiological measurement. *Physiological Measurement*, *28*(3), R1–R39. doi:10.1088/0967-3334/28/3/R01

Andreassi, J. L. (2006). *Psychophysiology: Human behavior and physiological response* (Fifth ed.). Mahwah, NJ: Lawrence Erlbaum Associates.

Ball, J. (2002). Lost. On *Prairie Runner* [CD]. Arvada, CO: Red Feather Music.

Berman, M. H. (2001, November 28). *EEG biofeedback training for children seen in the public mental health system with attention deficit hyperactivity disorder* (Quietmind Foundation internal report). Retrieved from the Quietmind Foundation website: http://quietmindfdn.org/docs/OMHreport.doc

Berntson, G. G., Bigger, J. T., Jr., Eckberg, D. W., Grossman, P., Kaufmann, P. G., Malik, M., … van der Molen, M. W. (1997). Heart rate variability: Origins, methods, and interpretive caveats. *Psychophysiology*, *34*(6), 623–648. doi:10.1111/j.1469-8986.1997.tb02140.x

Black Hawk, & J. B. Patterson, J. B. (Ed.) (1834). *Life of Ma-ka-tai-me-she-kia-kiak or Black Hawk – Dictated by Himself*. Boston, MA: Russell, Odiorne, & Metcalf, Boston.

Burton, F. R. (1909). *American Primitive Music – with Especial Attention to the Songs of the Ojibways*. New York, NY: Moffat, Yard and Company, New York.

Darling, D. (2009). Prayer for compassion. On *Prayer for Compassion* [CD]. Boulder, CO: Wind Over the Earth Records.

Deloria, E. C., & Jay Brandon (tr.) (1961, June). The origin of the courting flute, a legend in the Santee Dakota dialect. *University of South Dakota, Museum News*, *22*(6), 1–7.

Densmore, F. (1918). *Teton Sioux Music and Culture*. Smithsonian Institution, Bureau of American Ethnology, Bulletin 61. Washington, DC: United States Government Printing Office.

Densmore, F. (1923). *Mandan and Hidatsa music* (Smithsonian Institution, Bureau of American Ethnology, Bulletin 80). Washington, DC: Government Printing Office.

Densmore, F. (1929). *Chippewa customs* (Smithsonian Institution, Bureau of American Ethnology, Bulletin 86). Washington, DC: Government Printing Office.

Densmore, F. (1936). *The American Indians and their music* (Rev. ed.). New York, NY: The Woman's Press.

Densmore, F. (1957). *Music of Acoma, Isleta, Cochiti, and Zuñi pueblos* (Smithsonian Institution, Bureau of American Ethnology, Bulletin 165). Washington, DC: Government Printing Office.

Dubé, R. A. (2007). *Songs of the spirit – Attending to aboriginal students' emotional and spiritual needs through a Native American flute curriculum* (M.Ed. dissertation, University of Saskatchewan, Saskatoon, Canada). Retrieved from the University of Saskatchewan Library website: http://library.usask.ca/

Eley, R., & Gorman, D. (2008). Music therapy to manage asthma. *Aboriginal and Islander Health Worker Journal*, *32*(1), 9–10.

Elliot, W., Izzo, J., White, W., Rosing, D., Snyder, C., Alter, A., … Black, H. R. (2004). Graded blood pressure reduction in hypertensive outpatients associated with use of a device to assist with slow breathing. *Journal of Clinical Hypertension*, *6*, 553–559.

Erdoes, R. (transcriptions and Ed.), & Goble, P. (pictures) (1976). *The sound of flutes and other Indian legends – told by Lame Deer, Jenny Leading Cloud, Leonard Crow Dog, and others*. New York, NY: Pantheon Books.

Erdoes, R., & Ortiz, A. (1984). *American Indian Myths and Legends*. New York, NY: Pantheon Books.

Friedman, B. H. (2007). An autonomic flexibility-neurovisceral integration model of anxiety and cardiac vagal tone. *Biological Psychology*, 74(2), 185–199.

Giardino, N., Chan, L., & Borson, S. (2004). Combined heart rate variability and pulse oximetry biofeedback for chronic obstructive pulmonary disease: Preliminary findings. *Applied Psychophysiology & Biofeedback*, *29*, 121–133.

Gilman, B. I. (1908). *Hopi songs* (A Journal of American Ethnology and Archæology, Vol. 5). Boston, MA: Houghton Mifflin Company.

Goldberger, A. L. (1991). Is the normal heartbeat chaotic or homeostatic? *News in Physiological Science*, *6*, 87–91.

Goldberger, A. L., Rigney, D. R., Mietus, J., Antman E. M., & Greenwald, S. (1988). Nonlinear dynamics in sudden cardiac death syndrome: Heartrate oscillations and bifurcations. *Cellular and Molecular Life Sciences*, *44*(11–12), 983–987. doi:10.1007/BF01939894

Goncharova, I. I., McFarland, D. J., Vaughan, T. M., & Wolpaw, J. R. (2003). EMG contamination of EEG: spectral and topographical characteristics. *Clinical Neurophysiology*, *114*(9), 1580–1593. doi:10.1016/S1388-2457(03)00093-2

Goss, C. F. (2011). Anatomy of the Native American flute. Retrieved from the Flutopedia website: http://www.Flutopedia.com/anatomy.htm on April 12, 2012.

Goss. C. F. (2013). Intraoral pressure in ethnic wind instruments. arXiv:1308.5214. Retrieved August 25, 2013.

Goss, C. F., & Miller, E. B. (2013). Dynamic metrics of heart rate variability. arXiv:1308.6018. Retrieved August 30, 2013.

Gruzelier, J. (2009). A theory of alpha/theta neurofeedback, creative performance enhancement, long distance functional connectivity and psychological integration. *Cognitive Processes*, *10*, Supplement 1, S101–S109. doi:10.1007/s10339-008-0248-5

Gruzelier, J., & Egner, T. (2005). Critical validation studies of neurofeedback. *Child and Adolescent Psychiatric Clinics of North America*, *14*, 83–104. doi:10.1016/j.chc.2004.07.002

Gumilla, J. (1741). *El Orinoco ilustrado y defendido* [The Orinoco illustrated and defended]. Madrid: Manuel Fernandez.

Harmat, L., & Theorell, T. (2010). Heart rate variability during singing and flute playing. *Music and Medicine*, *2*, 10–17. doi:10.1177/1943862109354598

Harmat, L., Ullén, F., de Manzano, Ö., Olsson, E., Elofsson, U., von Schéele, B., & Theorell, T. (2011). Heart rate variability during piano playing: A case study of three professional solo pianists playing a self-selected and a difficult prima vista piece. *Music and Medicine*, *3*(2), 102–107. doi:10.1177/1943862110387158

Henriques, G., Keffer, S., Abrahamson, C., & Horst, S. J. (2011). Exploring the effectiveness of a computer-based heart rate variability biofeedback program in reducing anxiety in college students. *Applied Psychophysiology & Biofeedback*, *36*(2), 101–112.

Hill, J. D., & Chaumeil, J.-P. (Eds.) (2011). *Burst of breath: Indigenous ritual wind instruments in lowland South America*. Lincoln, NE: University of Nebraska Press.

Hornbostel, E. M. von, & Sachs, C. (1914) . Abhandlungen und Vorträge - Systematik der Musikinstrumente [Essays and lectures - classification of musical instruments]. *Zeitschrift für Ethnologie – Organ der Berliner Gesellschaft für Anthropologie, Ethnologie und Urgeschiechte*, *46*, 553–590.

Hough, W. (1918). The Hopi Indian collection in the United States National Museum. *Proceedings of the United States National Museum*, *54*, 235–296.

Jensen, O., Goel, P., Kopell, N., Pohja, M., Han, R., & Ermentrout, B. (2005). On the human sensorimotor-cortex beta rhythm: Sources and modeling. *NeuroImage*, *26*(2), 347–355. doi:10.1016/j.neuroimage.2005.02.008

Jones, M. J. (2010). *Revival and community: The history and practices of a Native American flute circle* (M.A. dissertation, Kent State University). Retrieved from the OhioLink ETD Center website: http://rave.ohiolink.edu/

Karavidas, M. (2008). Heart rate variability biofeedback for major depression. *Biofeedback*, *36*(1), 18–21.

Karavidas, M. K., Lehrer, P. M., Vaschillo, E., Vaschillo, B., Marin, H., Buyske, S., … Hassett, A. (2007). Preliminary results of an open label study of heart rate variability biofeedback for the treatment of major depression. *Applied Psychophysiology & Biofeedback*, *32*, 19–30. doi:10.1007/s10484-006-9029-z

Komárková, Z. (2012). *Rehabilitace chronických respiračních poruch formou hry na dechové nástroje* [Rehabilitation of chronic respiratory problems by playing a wind instruments] (Bachelor's dissertation, Univerzita Palackého in Olomouci). Retrieved from the Theses.cz website: http://theses.cz/.

Krispijn, T. J. H. (2008). Music and healing for someone far away from home: HS 1556, a remarkable Ur III incantation, revisited. In R. J. van der Spek (Ed.), *Studies in Ancient Near Eastern World View and Society* (pp. 173–194). Bethesda, MD: CDL Press.

Lehrer, P., Carr, R., Smetankine, A., Vaschillo, E., Peper, E., Porges, S., … Hochron, S. (1997). Respiratory sinus arrhythmia versus neck/trapezius EMG and incentive inspirometry biofeedback for asthma: A pilot study. *Applied Psychophysiology & Biofeedback*, *22*, 95–109.

Lehrer, P., Vaschillo, E., Vaschillo, B., Lu, S., Scardella, A., Siddique, M., & Habib, R. H. (2004). Biofeedback treatment for asthma. *Chest*, *126*, 352–361.

Lehrer, P. M., Vaschillo, E., & Vaschillo, B. (2000). Resonant frequency biofeedback training to increase cardiac variability: Rationale and manual for training. *Applied Psychophysiology and Biofeedback*, Vol. *25*(3), 177–191.

Lindh, W., Pooler, M., Tamparo, C., & Dahl, B. M. (2009). *Delmar's Comprehensive Medical Assisting: Administrative and Clinical Competencies* (Fourth ed.). Clifton Park, NY: Cengage Learning.

Lubar, J. F., & Shouse, M. N. (1976). EEG and behavioral changes in a hyperkinetic child concurrent with training of the sensorimotor rhythm (SMR) – A preliminary report. *Biofeedback and Self-Regulation*, *1*(3), 293–306.

Lucia, R. (1994). Effects of Playing a Musical Wind Instrument in Asthmatic Teenagers. *Journal of Asthma*, *31*(5), 375–385. doi:10.3109/02770909409061317

Lucking, C. H., Creutzfeldt, O. D., Heinemann, U. (1970). Visual evoked potentials of patients with epilepsy and of a control group. *Electroencephalography and Clinical Neurophysiology*, *29*, 557–566. doi:10.1016/0013-4694(70)90098-2

Marks, M. B. (1974). Musical wind instruments in rehabilitation of asthmatic children. *Annals of Allergy*, *33*(6), 313–319.

Mendes, W. B. (2009). Assessing autonomic nervous system activity. In E. Harmon-Jones & J. Beer (Eds.), *Methods in Social Neuroscience* (pp. 118–147). New York, NY: Guilford Press.

Menezes Bastos, R. J., & Rodgers, D. A. (trans.) (2007). Música nas sociedades indígenas das terras baixas da América do Sul: estado da arte [Music in the indigenous societies of lowland South America: the state of the art]. *Mana*, *13*(2), 293–316. English translation by D. A. Rodgers retrieved from the SciELO website: http://SocialSciences.scielo.org/

Metzger, L. K. (2006). An existential perspective of body beliefs and health assessment. *Journal of Religion and Health*, *45*(1), 130–146. doi:10.1007/s10943-005-9008-3

Meymandi, A. (2009). Music, Medicine, Healing, and the Genome Project. *Psychiatry* (Edgemont), *6*(9), 43–45.

Miller, E. B. (2011). *Bio-guided music therapy: A practitioner's guide to the clinical integration of music and biofeedback*. London, England: Jessica Kingsley.

Monastra, V. J., Lynn, S., Linden, M., Lubar, J. F., Gruzelier, J., LaVaque, T. J. (2005). Electroencephalographic biofeedback in the treatment of attention-deficit/hyperactivity disorder. *Applied Psychophysiology and Biofeedback*, 30(2), 95–114. doi:10.1007/s10484-005-4305-x

Moss, D. (2004). Heart rate variability and biofeedback. *Psychophysiology Today: The Magazine for Mind-Body Medicine*, *1*, 4–11.

Nahshoni, E., Aravot, D., Aizenberg, D., Sigler, M., Zalsman, G., Strasberg, B., … Weizman, A. (2004). Heart rate variability in patients with major depression. *Psychosomatics*, *45*, 129–134.

Nakahara, H., Furuya, S., Obata, S., Masuko, T., & Kinoshita, H. (2009). Emotion-related changes in heart rate and its variability during performance and perception of music. *Annals of the New York Academy of Sciences*, *1169*, 359–362. doi:10.1111/j.1749-6632.2009.04788.x

Nakahara, H., Furuya, S., Francis, P. R., & Kinoshita, H. (2010). Psycho-physiological responses to expressive piano performance. *International Journal of Psychophysiology*, *75*, 268–276. doi:10.1016/j.ijpsycho.2009.12.008x

Nakai, R. C. (1993). Canyon people. On *Canyon Trilogy* [CD]. Phoenix, AZ: Canyon Records.

Petrů, V., Carbolová, A., & Kloc, V. (1993). Zobcová flétna jako pomůcka při léčbě respiračních onemocnění [The recorder as a treatment aid in respiratory diseases]. *Ceskoslovenska Pediatrie*, *48*(7), 441–442.

Piedade, A. T. de C. (2004). *O Canto do Kawoká: Música, Cosmologia e Filosofia Entre os Wauja do Alto Xingu* [The song of Kawoká: Music, cosmology and philosophy among the Upper Xingu Wauja] (Doctoral dissertation, Universidade Federal de Santa Catarina, Florianópolis, Brazil).

Piedade, A. T. de C. (2006). Reflexões a partir da etnografia da música dos índios Wauja [Reflections from the ethnography of music of the Wauja Indians]. *Revista Anthropológicas*, *17*(1), 35–48.

Puhan, M. A., Suarez, A., Lo Cascio, C., Zahn, A., Heitz, M., & Braendli, O. (2006). Didgeridoo playing as alternative treatment for obstructive sleep apnoea syndrome: Randomised controlled trial. *British Medical Journal*, 332:266. doi:10.1136/bmj.38705.470590.55

Renaud, E.-B. (1926). Flûtes indiennes préhistoriques du Sud-Ouest Américain [Indian flutes of prehistoric American Southwest]. *Bulletin de la Société préhistorique de France*, *23*(7–8), 168–178. doi:10.3406/bspf.1926.5910

Russoniello, C. V., Zhirnov, Y. N., Pougatchev, V. I., & Gribkov, E. N. (2013). Heart rate variability and biological age: Implications for health and gaming. *Cyberpsychology, Behavior, and Social Networking*, *16*(4), 302–308. doi:10.1089/cyber.2013.1505

Sandercock, G. H., Bromley, P. D., & Brodie, D. A. (2005). Effects of exercise on heart rate variability: Inferences from meta-analysis. *Medicine and Science in Sports and Exercise*, *37*(3), 433–439. doi:10.1249/01.MSS.0000155388.39002.9D

Selvaraj, N., Jaryal, A., Santhosh, J., Deepak, K. K., & Anand, S. (2008). Assessment of heart rate variability derived from finger-tip photoplethysmography as compared to electrocardiography. Journal of Medical Engineering and Technology, *32*(6), 479–484. doi:10.1080/03091900701781317

Shackman, A. J., McMenamin, B. W., Slagter, H. A., Maxwell, J. S., Greischar, L. L., & Davidson, R. J. (2009). Electromyogenic artifacts and electroencephalographic inferences. *Brain Topography*, *22*(1), 7–12. doi:10.1007/s10548-009-0079-4

Silber, M. H., Ancoli-Israel, S., Bonnet, M. H., Chokroverty, S., Grigg-Damberger, M. M., Hirshkowitz, M., … Iber, C. (2007). The visual scoring of sleep in adults. *Journal of Clinical Sleep Medicine*, *3*(2), 121–131.

Sowder, E., Gevirtz, R., Shapiro, W., & Ebert, C. (2010). Restoration of vagal tone: A possible mechanism for functional abdominal pain. *Applied Psychophysiology and Biofeedback*, 35(3), 199–206. doi:10.1007/s10484-010-9128-8

Stacey, R. (1906). Some Zuni ceremonies and melodies. *The Musiclovers Calendar*, *2*(1), 54–61.

Task Force of the European Society of Cardiology and the North American Society of Pacing and Electrophysiology (1996). Heart rate variability: Standards of measurements, physiological interpretation, and clinical use. *European Heart Journal*, *17*, 354–381.

Thatcher, R. W. (2008). *NeuroGuide Manual and Tutorial*. St. Petersburg, FL: Applied Neuroscience. Retrieved from the Applied Neuroscience website: http://www.AppliedNeuroscience.com/NeuroGuide_Deluxe.pdf

Thurber, M. R. (2006). *Effects of heart-rate variability biofeedback training and emotional regulation on music performance anxiety in university students* (Doctoral dissertation, University of North Texas). Retrieved from the University of North Texas Digital Library website: http://digital.library.unt.edu/

van Deursen, J. A., Vuurman, E. F., Verhey, F. R., van Kranen-Mastenbroek, & Riedel, W. J. (2008). Increased EEG gamma band activity in Alzheimer's disease and mild cognitive impairment. *Journal of Neural Transmission*, *115*(9), 1301–1311. doi:10.1007/s00702-008-0083-y

Vaschillo, E. G., Vaschillo, B., & Lehrer, P. M. (2006). Characteristics of resonance in heart rate variability stimulated by biofeedback. *Applied Psychophysiology and Biofeedback*, *31*(2), 129–142. doi:10.1007/s10484-006-9009-3

Vernon, D. J. (2005). Can neurofeedback training enhance performance? An evaluation of the evidence with implications for future research. Applied Psychophysiology and Biofeedback, 30(4). doi:10.1007/s10484-005-8421-4

Wheat, A. L., & Larkin, K. T. (2010). Biofeedback of heart rate variability and related physiology: A critical review. *Applied Phychophysiology and Biofeedback*, *35*(3), 229–242. doi:10.1007/s10484-010-9133-y

Whitham, E. M., Lewis, T., Pope, K. J., Fitzgibbon, S. P., Clark, C. R., Loveless, S., … Willoughby, J. O. (2008). Thinking activated EMG in scalp electrical recordings. *Clinical Neurophysiology*, *119*(5), 1166–1175. doi:10.1016/j.clinph.2008.01.024

Wiand, L. L. (2001). *The effects of a sacred/shamanic music on trauma related disorders: Dissociative disorders and music of an indigenous Native American flute* (Doctoral dissertation, University of Detroit Mercy). Copy provided courtesy of the author.

Wiand, L. L. (2006). The effects of sacred/shamanic flute music on trauma and states of consciousness. *Subtle Energies and Energy Medicine*, *17*(3), 249–284.

Winship, G. P. (1896). The Coronado expedition, 1540–1542. In J. W. Powell (Dir.), *Fourteenth Annual Report of the Bureau of American Ethnology to the Secretary of the Smithsonian Institution, 1892-93* (Part 1), pp. 329–613.

Wissler, C. (1905). The whirlwind and the elk in the mythology of the Dakota. *The Journal of American Folk-lore*, *18*(71), 257–268.

Wright, L. D. (2006). Meditation: A new role for an old friend. *American Journal of Hospice & Palliative Medicine*, *23*(4), 323–327. doi:10.1177/1049909106289101

Xiangcai, X. (2000). Qigong for Treating Common Ailments: The Essential Guide to Self Healing. Boston, MA: YMAA Publication Center.

Yang, Z.-X., Cai, X., Liu, X.-Y. Qin, J. (2008). Relationship among eye condition sensitivities, photosensitivity and epileptic syndromes. *Chinese Medical Journal*, *121*(17), 1633–1637.

Young, R. W. (1939). Terminology for logarithmic frequency units. *Journal of the Acoustical Society of America*, *11*(1), 134–139. doi:10.1121/1.1916017

Žilka, V. (1993). *Veselé pískání – zdravé dýchání* [Merry whistling – healthy breathing]. Prague, Czech Republic: Panton, 1993.

Zucker, T. L., Samuelson, K. W., Muench, F., Greenberg, M. A., & Gevirtz, R. N. (2009). The effects of respiratory sinus arrhythmia biofeedback on heart rate variability and posttraumatic stress disorder symptoms: A pilot study. *Applied Psychophysiology & Biofeedback*, *34*(2), 135–143. doi:10.1007/s10484-009-9085-2

Eric B. Miller is now Adjunct Professor and Coordinator of the David Ott Laboratory for Music and Health in the Department of Music Therapy at Montclair State University.

We would like to acknowledge the flute players at the Flute Haven Native Flute School who participated in this study. Note that both authors receive an honorarium for instruction at the Flute Haven seminar as well as facilitation in other music instruction and therapy contexts.

Correspondence should be sent to Dr. Eric B. Miller, John J. Cali School of Music, Montclair State University, 1 Normal Ave., Montclair, NJ 07043, Phone: 610-999-5960, E-mail: millerer@montclair.edu and Dr. Clinton F. Goss, 6 Fieldcrest Road, Westport, CT 06880, Phone: 203-454-1479, E-mail: clint@goss.com.

Dr. Eric Miller, Ph.D. MT-BC biofeedback therapist and board-certified music therapist is author of *Bio-guided Music Therapy* (2011), Jessica Kingsley Publishers. He serves as Coordinator for the Ott Lab for Music & Health at Montclair University in the Music Therapy Dept. Recent international presentations include sessions at the World Music Therapy Congress in Seoul, S. Korea, Hsien Chuan University in Tapei, Taiwan, and workshops in Kiental, Switzerland, Cogolin, France and Torino, Italy. Dr. Miller serves as Executive Director of nonprofits *Music for People*, *Expressive Therapy Concepts* and founded the *Biofeedback Network*. Miller collaborated with Grammy-winning cellist, David Darling on the instrumental CD *Jazzgrass*.

Clinton F. Goss, Ph.D. is a computer scientist and music facilitator. He has worked as a consultant in software development and Internet technologies for 36 years, and has recently served as expert witness in U.S. federal and international software intellectual property cases. Since 1998, he and his wife, Vera, have engaged in technical assistance projects in developing countries, visiting 86 countries. He is also a musician on an array of world instruments, primarily ethnic flutes, and is actively engaged in music facilitation. His areas of publication include compiler optimization, graph theory, quantitative metrics of physiology, and music facilitation, as well as over 20 CDs under his *Manifest Spirit* label. He holds an FAA flight instructor certificate, a certificate in music facilitation from the *Music for People* organization, and a Ph.D. in computer science from Courant Institute, New York University.